# Improved Multi-GPU parallelization of a Lagrangian Transport Model

Saheed Bolarinwa
Computational Neuroscience
Eotvos Lotrand university,
Budapest
Hungary
bolarinwa@student.elte.hu

**Abstract** *Massive-Parallel Trajectory Calculations (MPTRAC) is a Lagrangian Transport Model developed by the Jülich Supercomputing Centre. It is open-sourced and available at* `https://github.com/slcs-jsc/mptrac`. *The current version 2.2 of MPTRAC adds support for parallelization on Graphics Processing Unit (GPU). It uses the MPI-OpenMP-OpenACC threefold heterogeneous parallelization. This report highlights our work on improving GPU parallelization by supporting compute nodes with multiple GPUs. However, since the default support for multi-GPUs in OpenACC is limited[6], the current implementation allows each MPI process to access only a single GPU. Thus, the only way to take full advantage of multi-GPU nodes in the current version is to launch multiple processes, which increases resource contention. We investigated the benefits of having only one process offload to all available GPU devices. Our evaluation shows that our implementation is beneficial to MPTRAC, not just in terms of performance but also in flexibility. With our multi-GPU version, we achieve over three times speedup for some MPTRAC modules. We have made MPTRAC more flexible so that users can decide on the number of GPUs each MPI process can access. Depending on the cluster's configuration, users may choose differently.*

# 1 Introduction

## 1.1 What is a Lagrangian Transport Model?

Atmospheric transport highlights the path or trajectory through which various pollutants travel from the source to other continents, both coastal and open sea[1]. Modelling such complex transport is a frequent practice in many fields of geosciences. Scientists study these models to answer questions relating to hydro-metrology, air quality, greenhouse gasses, responses to volcanic eruptions and nuclear releases[4].

Lagrangian modelling is an approach to modelling complex atmospheric transport. Another approach is the Eulerian model, which uses fixed spatial grids and transport. It is based on the fluid flows between the grid cells[3].

Lagrangian transport models simulate the transport of irregularly distributed trace gas and aerosols as computational particles and track their trajectories and the processes acting on them. One advantage of Lagrangian models is that they can simulate even tiny features, which often appear as elongated air filaments in the tropospheric and stratospheric flows[3].

Air pollution, nuclear accidents, volcanic eruptions, wildfires, and assessing the circulation of the troposphere and stratosphere are some areas where Lagrangian Transport Models have been used for practical applications from a local to a global scale. Hoffmann et al.[3] expounds on use cases of Lagrangian Transport Models and various Lagrangian Transport Models that exist, both for research and operational purpose.

## 1.2 What is MPTRAC[5]?

Massive-Parallel Trajectory Calculations (MPTRAC) is a Lagrangian Transport Model developed at the Jülich Supercomputing Centre. Hoffmann et al.[3] introduced it as a model for Lagrangian transport simulations of volcanic emissions in the free troposphere and stratosphere. It is written in C language and open-sourced at https://github.com/slcs-jsc/mptrac. Version 2.2 of MPTRAC began the quest to exploit the potential for accelerating MPTRAC on Graphics Processing Units (GPUs). It uses a threefold hybrid parallelization involving Message Passing Interface (MPI), Open Multi-Processing (OpenMP) and OpenACC. In previous evaluations[2], it delivered better performance over the CPU version.

## 1.3 MPI-OpenMP-OpenACC threefold hybrid Parallelisation.

***Message Passing Interface (MPI)*** is a standardized interface and communication protocol for programming high-performance parallel computers. It is a specification designed to be portable and scalable. The interface primarily allows point-to-point and collective communication (Message Passing) between computing tasks on participating nodes on the cluster. There are several implementations of the MPI standard in multiple programming languages. In MPTRAC, MPI is used for cluster-level parallelization.

***Open Multi-Processing (OpenMP)*** is a programming interface for multi-platform parallel programming on shared-memory multiprocessing units. It is implemented as a compiler extension to various programming languages on multiple architectures. OpenMP primarily distributes the workload over a set of computing threads on a compute node. The interface consists of compiler directives (pragmas), library routines, and environment variables. In MPTRAC, OpenMP is used for node-level CPU parallelization.

***Open accelerators (OpenACC)*** is also a programming interface specifically targeted at accelerators like GPUs. Accelerators like GPUs create a heterogeneous environment which is often challenging to program. OpenACC aims to ease the parallel programming of heterogeneous CPU-GPU systems. Its interface is similar to the OpenMP interface as they are both based on compiler directives (pragmas), library routines, and environment variables. In MPTRAC, OpenACC is used for node-level GPU parallelization.

Version 2.2 of MPTRAC introduces a threefold hybrid parallelization that involves MPI, OpenMP and OpenACC. Section 3 expounds on the code structure and parallelization strategy. The improvements made to the current strategy by adding support for multi-GPUs are discussed in Section 4 while Section 5 highlights the results of the evaluation of the current improvement.

## 2 Problem Description

Hoffmann et al.[3] reported performance improvements on the GPU version of MPTRAC. However, since the default support for multi-GPUs in OpenACC is limited[6], the current implementation allows each process to offload to only a single GPU, notwithstanding the number of GPUs on the node. Figure 5.1 shows a case when only a single MPTRAC process runs on a node with four GPUs. In this situation, only one out of the four GPUs is in use others are completely idle. The kernels like "*Module Advection*" run in serial order on the only GPU available to the process. With the current version of MPTRAC, to take full advantage of multi-GPU nodes, the only option is to launch multiple MPI processes. Thus increasing the contention for the system resources. It is desirable to explore the possibility of having only a single process utilise all the available GPUs on the node, not just one. One of the areas we expect to see an improvement is in the output section of the code since the single process only needs to wait for the GPUs, not other MPI processes.

The goal of this project is to implement multi-GPU support in MPTRAC.

# 3 MPTRAC'S CODE STRUCTURE

MPTRAC is realeased as open sourced software. It is hosted at https://github.com/slcs-jsc/mptrac. The codebase is structured into three sections: the Input, the Output and the Chemistry-Physics Processing sections. Figure 3.1 shows an edited version of MPTRAC's call graph.

## 3.1 THE INPUT SECTION

Since all the data is stored on file, the input section involves reading from raw files into the system's memory. The netCDF[7] library is used for file-IO. The three main input functions are:

- `read_ctl()`: this function reads the model control parameters from the file.

- `read_atm()`: this function reads the initial particle data from the file.

- `read_met()` : this function reads various meteorological input data from multiple files. It makes further calls to functions that deal directly with reading specific meteorological data files ( `read_met_grid()`, `read_met_levels()`, and `read_met_surface()`). The function also perfoms extrapolation and sampling of the meteorological data ( `read_met_extrapolate()`, `read_met_periodic()`, `read_met_sample()`, and `read_met_detrend()`), and calculates additional meteorological variables ( `read_met_geopot()`, `read_met_pv()`, `read_met_pbl()`, `read_met_tropo()`, `read_met_cloud`, and `read_met_cape()`). The MPTRAC model provides the flexibility of using different meteorological data by allowing users to recalculate the values from different meteorological data sources.

## 3.2 THE CHEMISTRY-PHYSICS PROCESSING SECTION

For each air parcel in the large atmospheric dataset a couple of module operations are performed. Some of these modules are:

- `module_advection()`: calculate kinematic trajectories of the particles using given ⃗velocities.

- `module_diffusion_turb()`: adds stochastic perturbations to the trajectories to simulate the effects of turbulent diffusion

- `module_diffusion_meso()`: adds stochastic perturbations to the trajectories to simulate subgrid-scale wind fluctuations

- `module_convection()` and `module_sedi()`: alter the particle positions along the trajectories by simulating the effects of unresolved convection and sedimentation.

- `module_isosurf()`: constrains the particle positions to different types of isosurfaces.

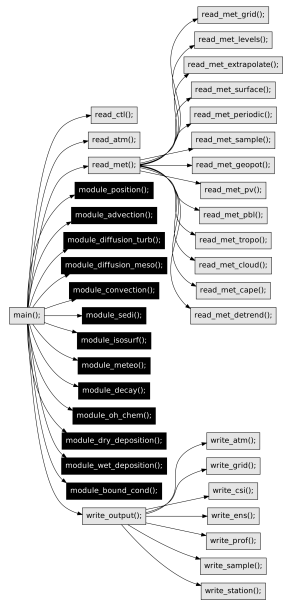

Figure 3.1: MPTRAC's call graph show only the key functions. Source: Hoffmann et al. [2]

- module_position(): enforces the particle to remain within the boundaries of the model domain.
- module_meteo() allows us to sample the meteorological data along the trajectories at fixed time intervals.

## 3.3 THE OUTPUT SECTION

The model output is directed via a generic function `write_output()` towards individual functions that can be used to write particle data `write_atm()`, gridded data `write_grid()`, or ensemble data for groups of particles `write_ens()`[2].

## 3.4 PARALLELIZATION STRATEGY

In the Lagrangian transport model, trajectory simulations of each air parcel are done independently. As a result, there is no communication between the kernels, making the workload "embarrassingly parallel" and thus suitable for GPUs. The previous versions of MPTRAC employ CPU-only parallelization, implemented with OpenMP. While MPI implementation enables cluster-level parallelization. In the current version, OpenACC is used to handle GPU offloading.

An advantage of using OpenACC is that both the code for CPU parallelization written in OpenMP and the OpenACC code for GPU parallelization coexist in the same source file. The existence of OpenACC indicates that there is a GPU, and this is checked with the `#ifdef` directive. For Instance:

LISTING 1: OPENMP AND OPENACC IN THE SAME CODEBASE

```
#ifdef _OPENACC
...

OpenACC code here
...

```

```
 7  #else
 8  ...
 9
10  OpenMP code here
11  ...
12
13  #endif
```

This strategy, however places some restrictions on the OpenACC implementation. For instance, instead of having to separately create, update and delete variables within the data region on the GPU, a single data clause like `copy`, `copyin`, `copyout` would have been enough. The problem is that this would require the scope to be defined and enclosed with braces. Since the code within that braces is offloaded to the GPU, we would have to duplicate the logic for OpenMP acceleration.

## 4 Multi-GPU Parallelization

The essence of adding multi-GPU support to OpenACC code is looping over the existing devices and setting the current device within each loop. There are at least two ways to implement this:

1. Normal for-loop:

**Listing 2:** A for-loop over all GPUs on the node

```
1   for(int device_num = 0; device_num < num_devices; ++device_num) {
2       acc_set_device_num(device_num, acc_device_nvidia);
3
4       ...
5       // OpenACC pragmas and GPU related logic
6       ...
7   }
```

2. OpenMP loop:

**Listing 3:** An OpenMP-loop over all GPUs on the node

```
1   #pragma omp parallel num_threads(num_devices)
2   {
3       int device_num = omp_get_thread_num();
4       acc_set_device_num(device_num, acc_device_nvidia);
5
6       ...
7       // OpenACC pragmas and GPU related logic
8       ...
9   }
```

The for-loop version would run on a single CPU thread, thus offloading to the GPU serially. Though we now have access to all GPUs, only one GPU is active at any moment. The OpenMP loop, on the other hand is a parallel loop running in parallel on different OpenMP threads. Thus, the GPU offloading in this case would run in parallel.

However, using the OpenMP loop requires considering that the OpenMP thread ordering is nondeterministic. Also, the use of the MPI-OpenMP-OpenACC hybrid parallelisation strategy calls for caution to avoid undesired nesting of OpenMP loops. Another issue that was considered is the diagnostic timing measurement obtained via the NVIDIA Tools Extension Library (NVTX).

Beyond these considerations, most of the work on loop parallelization has been done in MPTRAC version 2.2[2] The major aspect of multi-GPU porting is the data movement strategy. Most data structures have been designed and optimized for a single GPU. Below are some key areas focused on while porting MPTRAC for multi-GPU.

## 4.1 main() function:

***Initialising all GPUs:*** The existing implementation of MPTRAC initialises the GPU in the main function. To port this to multi-GPU, the code is wrapped inside a loop over all available GPUs as shown in Listing 4.

**LISTING 4:** INITIALIZING ALL GPUS

```
for(int device_num = 0; device_num < num_devices; device_num++) {
   acc_set_device_num(device_num, acc_device_nvidia);

   SELECT_TIMER("ACC_INIT", "INIT", NVTX_GPU);
   acc_device_t device_type = acc_get_device_type();
   acc_init(device_type);
}
```

***Creating data regions on all GPUs:*** is similar to Listing 4 except that in this case, within the loop over the GPU devices, we are creating a data region on all GPUs with:

```
#pragma acc enter data create
```

***Updating data regions on all GPUs:*** The GPU kernels operate on various kinds of geophysical data, all of which are stored in raw data files. The host extracts and populates the respective host variables. For most data, their whole is needed by the kernels, however, the atmospheric data needs to be distributed among the GPUs. It is also the output of the kernel, so the host will have to merge the atmospheric data obtained from each GPU. The algorithm is such that each GPU can independently work on a section of the data. While it is possible to split the atmospheric data that is transferred to the GPU, to simplify the implementation, all of it is sent to each GPU. A function, `calc_device_workload_range()` was created which allocates a portion of the data to each GPU and returns the range. This range is used to determine the range of the loop in which the various modules are called on the data.

To copy out the result, one of the approaches explored is the OpenACC deep copy, which is a relatively new feature in OpenACC. Subsection 4.4 elaborates more on the deep copy in OpenACC and the approach we used for the GPU-Host data transfer.

Since MPTRAC's strategy is to share the core logic between both OpenMP and OpenACC pragmas, the directives for updating the device variables are delayed until the host data has been fully populated from the data files. The code is similar to Listing 4 except that in this case, within the loop over the GPU devices, we are updating the listed variables within the data region on all GPUs with:

```
#pragma acc update device (atm, ..)
```

For the GPU to host direction:

```
#pragma acc update host (atm, ..)
```

***Deleting Data Regions on all GPUs:*** the code here is also similar to Listing 4. Here we use `#pragma acc wait` to ensure that the GPU is through with execution before deleting the data regions. Since we are placing this in a loop the directive is specific to the current device set by `acc_set_device_num(device_num, acc_device_nvidia)`.

The code looks like this:

**LISTING 5:** DELETING DATA REGIONS FROM ALL GPUS

```
for(int device_num = 0; device_num < num_devices; device_num++) {
    acc_set_device_num(device_num, acc_device_nvidia);

```

```
5       SELECT_TIMER("DELETE_DATA_REGION", "MEMORY", NVTX_GPU);
6   #pragma acc exit data delete(ctl,atm,cache,clim,met0,met1,dt,random_nums)
7   }
```

## 4.2 Extending the Climatological data structure type (clim_t)

Climatological data is loaded from a file in the function `read_clim()` and stored in an object of `struct clim_t`. Before operating on this data, MPTRAC sets an initial dataset for $HNO_3$ volume mixing ratios and tropopause pressure. The variables holding these datasets are global variables inside *libtrac.h*. This makes them easily accessible across several functions. However, implementing a multi-GPU data transfer for such global variables is challenging because the loop over multiple GPUs cannot exist out of context. It needs to be inside a function. There were a few solutions explored including creating these objects on the heap within the main function and passing around their pointer. All the options considered require some refactoring, however, incorporation of these variables into `struct clim_t` which contains other climatological data provided a cleaner implementation of the multi-GPU code. Since most of the affected functions call `read_clim()`, the data transfers are conducted inside it.

## 4.3 Random Number Generation on all GPUs

The modules which define the processes operating on each air parcel are called within the loop over timesteps. The time steps of each air parcel are also set inside this loop by calling `module_timesteps()`. Before calling any of the modules inside the loop a new series of random numbers is generated by calling `generate_random_nums()`. This is a new modification to the code which enables the decoupling of the random number generation logic from the algorithm expressed in the modules.

The random numbers are used in `module_diffusion_turb()`, `module_diffusion_meso()` and `module_convection()` to introduce varying degree of randomness to the calculations of some properties like wind perturbation, latitude, and longitude of atmospheric components. MPTRAC uses the NVIDIA CURAND library to generate random numbers. Each function has a different requirement for the distribution of the random numbers, so in the current version, they are generated within the functions using the same random number generator or seed. While this is okay for a single GPU, refactoring it for multi-GPU support involves duplication of codes within each function. This is what motivated the decoupling of the random number generation from the functions that use them.
In the recent version, a struct is defined as shown in Listing 6 and whenever `generate_random_nums()` is called within the timesteps loop, the object of the struct is updated with a new set of random numbers and passed to each function for usage. The seeding of the random generator is performed inside `module_rng_init()` which is where the random number generator is created. The NVIDIA CURAND API needs a random generator which it uses to generate random numbers on the current GPU. Unlike the current version where the random number generator is a pointer to a single object of `curandGenerator_t`, in the current version it points to as many objects as the number of devices available. So, `module_rng_init()` will generate a unique generator for each device, using the seed:
*num_of_mpi_task* + 83 ∗ *gpu_device_id*.

**Listing 6:** A STRUCT THAT HOLDS THE RANDOM NUMBERS GENERATED FOR EACH FUNCTION

```
1
2   /*! Random Numbers */
3   typedef struct {
```

```
4        double convection[NP]; /* Random numbers for calculating convection of air parcels */
5        double diff_meso[3*NP]; /* Random numbers for calculating mesoscale diffusion. */
6        double diff_turb[3*NP]; /* Random numbers for calculating turbulent diffusion. */
7    } randoms_t;
```

## 4.4 Moving atmospheric data using the OpenACC deepcopy feature

The deep copy problem is a data transfer problem which can occur even within the same physical memory. However, our case involved two physically separate memory units. It happens when we need to copy an object of a data structure that has pointers as its members from one memory context to another. The pointers inside the object point to a memory space with a local address. If we copy such an object to a GPU which uses a different physical address context, the pointers will then be pointing elsewhere based on the GPU's interpretation. To solve this problem, the objects pointed to must be copied with their pointers. We then obtain the memory address on the target GPU device and assign it to the corresponding pointer. This is known as deepcopying.

OpenACC has several features that make it easy to perform a deepcopy operation without much boilerplate code. Some are *attach*, *detach*, *shape* and *policy* clauses. It is also possible to implement manual copying.

The `atm_t` struct is a major data structure in MPTRAC. Most of the modules in MPTRAC operate on the atmospheric data stored in an object of struct `atm_t`. The object is updated and returned, so it is the actual output of the OpenACC kernels. The struct contains arrays of different atmospheric properties stored for each air parcel. The main concern in this project is the GPU to CPU transfer.

Though we explored various OpenACC deepcopy, our approach was to manually copy out only the range of workload for its GPU. As explained earlier this range is calculated inside the function `calc_device_workload_range()`. Since the ranges are unique to each GPU and do not overlap, the copy operation is guaranteed to avoid race condition errors. The OpenACC pragma looks like this:

```
1   #pragma acc update host(atm->np,atm->time[start:end],atm->p[start:end], \
2           atm->zeta[start:end],atm->lon[start:end],atm->lat[start:end], \
3           atm->q[0:NQ][start:end])
```

## 5 Results

We evaluated our multi-GPU version of MPTRAC on the JUWELS system at the Jülich Supercomputing Centre. The JUWELS Booster has 936 compute nodes, each equipped with four NVIDIA A100 Tensor Core GPUs.

Figure 5.1 shows the Nsight analysis of MPTRAC version 2.2 with NVIDIA Tools Extension (NVTX) timing measurements. Only one of the four GPUs available on each JUWELS Booster node is in use, as indicated by the single orange bar. In Figure 5.2, which is from the analysis of our multi-GPU version, the four orange bars indicate that four GPUs are in use. The effect of using multi-GPU is an average of 3.1 increase in speed. For instance, Module Advection which executed in an average of 113 ms on a single GPU within one process of MPTRAC v2.2, on our current version with four GPUs per process, executes in an average of 37 ms.

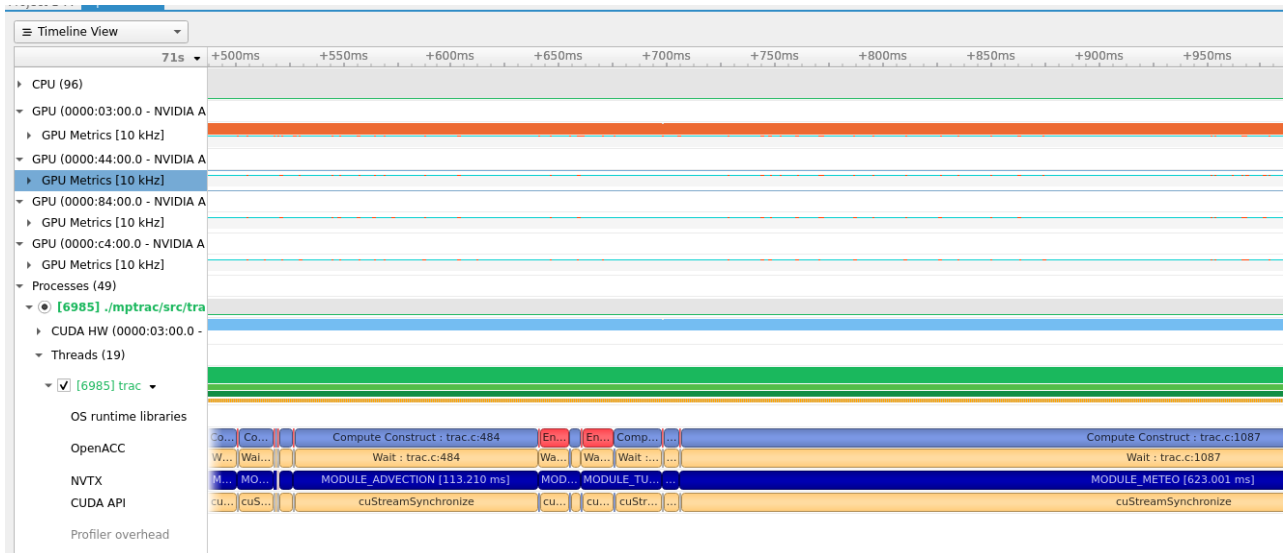

Figure 5.1: Nsight analysis of MPTRAC version 2.2 showing single GPU per each process. Module Advection and other are offload in serial

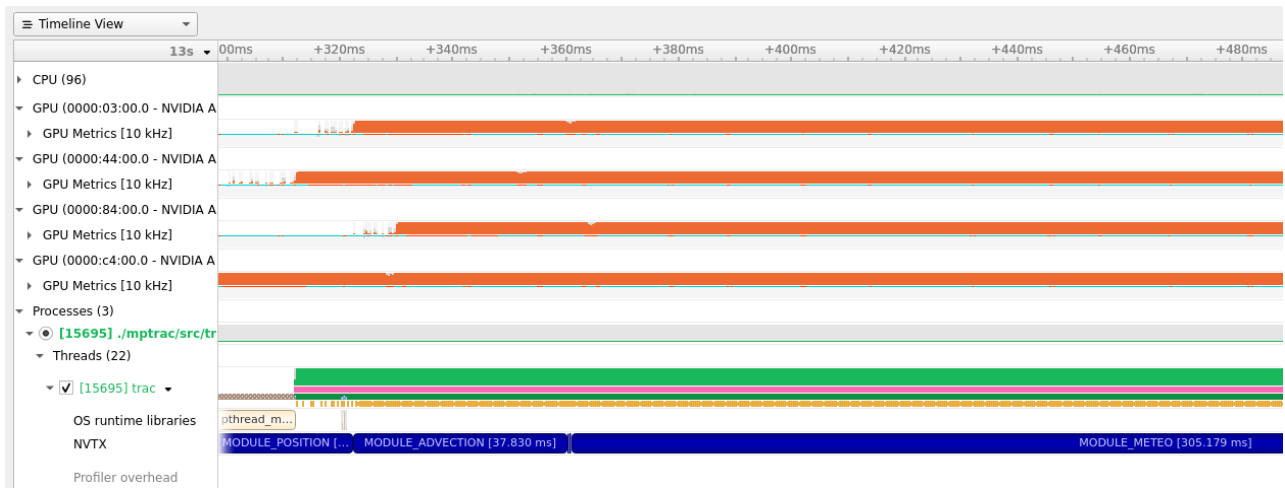

Figure 5.2: Nsight analysis of current multi-GPU showing multiple GPUs per each process. 4 Module Advection kernels are offloaded in parallel

# 6 Conclusion

With our version showing over three times speedup, we have demonstrated that MPTRAC can benefit from multi-GPU operations. We have also improved the flexibility of MPTRAC, making it possible to decide the number of GPUs to assign to the MPTRAC running process. This can be achieved by passing an integer to the MPTRAC binary. Any negative number is interpreted as "all available GPUs". We have been able to port a large part of MPTRAC for multi-GPU offloading. However, modules like those used for sorting are still pending. There are still a few open issues to be fixed in the output module, also the NVTX timing logic needs to be improved to properly support multi-GPU.

Our contributions lay a foundation for future improvements in the GPU support in MPTRAC.

# 7 Acknowledgments

I am grateful to my supervisors Dr Lars Hoffmann and Dr Kaveh Haghighi Mood. They didn't just supervice but actually collaborate and brainstormed with me on the project. I am also grateful to Dr Ivo Kabadshow who lead the coodination of the Guest Student Programme. I have learnt a lot during this programme. I am very grateful to all the staffs of the Juelich Supercomputing Center who made this possible.
Thank you all.